\documentclass[aps,prb,twocolumn, groupedaddress,superscriptaddress,showpacs]{revtex4-1}
\usepackage{graphicx,amsmath}
\usepackage{amssymb}
\usepackage{nicefrac}

\begin{document}
\title{Dynamical cluster approximation within an augmented plane-wave framework: Spectral properties of SrVO$_3$}

\author{Hunpyo Lee}
\affiliation{Institut f\"ur Theoretische Physik, Goethe-Universit\"at Frankfurt, Max-von-Laue-Stra{\ss}e 1, 60438 Frankfurt am Main, Germany}
\author{Kateryna Foyevtsova}
\affiliation{Institut f\"ur Theoretische Physik, Goethe-Universit\"at Frankfurt, Max-von-Laue-Stra{\ss}e 1, 60438 Frankfurt am Main, Germany}
\author{Johannes Ferber}
\affiliation{Institut f\"ur Theoretische Physik, Goethe-Universit\"at Frankfurt, Max-von-Laue-Stra{\ss}e 1, 60438 Frankfurt am Main, Germany}
\author{Markus Aichhorn}
\affiliation{Institute of Theoretical and Computational Physics, TU Graz, Petersgasse 16, Graz, Austria}
\author{Harald O. Jeschke}
\affiliation{Institut f\"ur Theoretische Physik, Goethe-Universit\"at Frankfurt, Max-von-Laue-Stra{\ss}e 1, 60438 Frankfurt am Main, Germany}
\author{Roser Valent\'\i}
\affiliation{Institut f\"ur Theoretische Physik, Goethe-Universit\"at Frankfurt, Max-von-Laue-Stra{\ss}e 1, 60438 Frankfurt am Main, Germany}

\newcommand{\svo}{SrVO$_3$}

\date{\today}

\begin{abstract}

We present a combination of local density approximation (LDA) with the
dynamical cluster approximation (LDA+DCA) in the framework of the
full-potential linear augmented plane-wave method, and compare our
LDA+DCA results for SrVO$_3$ to LDA with the dynamical mean field theory (LDA+DMFT) calculations as well as experimental observations
on SrVO$_3$. We find a qualitative agreement of the momentum resolved
spectral function with angle-resolved photoemission spectra (ARPES)
and former LDA+DMFT results. As a correction to LDA+DMFT, we observe
more pronounced coherent peaks below the Fermi level, as indicated by
ARPES experiments.  In addition, we resolve the spectral functions in the ${\bf
  K}_{0}=(0,0,0)$ and ${\bf K}_{1}=(\pi,\pi,\pi)$ sectors of DCA,
where band insulating and metallic phases coexist. Our approach can be
applied to correlated compounds where not only local quantum
fluctuations but also spatial fluctuations are important.
\end{abstract}

\pacs{71.10.Fd}

\keywords{}

\maketitle


\section{Introduction\label{Introduction}}
The development of reliable numerical tools for the description of the
electronic structure of correlated compounds is one of the most
challenging tasks in the condensed matter community. As an example,
transition-metal perovskites with partially filled $t_{2g}$ orbitals
are predicted to be conventional metals in the framework of
one-electron approaches like density functional theory (DFT) in the
local density approximation (LDA).  Nevertheless, a few perovskite
families show a markedly different behavior; SrVO$_3$ and CaVO$_3$ are
correlated metals with significant mass enhancement and LaTiO$_3$
displays features of a Mott
insulator~\cite{Imada1998,Pavarini2004}. In all compounds, this
anomalous behaviour may be caused by correlations resulting from
Coulomb repulsion effects. Therefore, progress on methods including
correlation effects beyond DFT is very desirable.

Dynamical mean field theory
(DMFT)~\cite{Metzner1989,Georges1996,Kotliar2004} takes local quantum
fluctuations fully into account but the momentum dependence of the
self-energy is neglected.  This method has been developed over the
last twenty years and successfully describes the metal to Mott
insulator transition in frustrated
systems~\cite{Liebsch2005,Koga2004,Inaba2005,Zitzler2004,Lee2010(3)}
and non Fermi-liquid behavior in multi-orbital
systems~\cite{Biermann2005,Werner2008,Ishida2010} to mention a few
examples.  On the other hand, it cannot describe such phases as spin
density wave or d-wave superconductivity due to its lack of spatial
correlations. In order to overcome these problems the first extension
to single-site DMFT are multi-site approaches in which the short-range
correlations are exactly considered within a
cluster~\cite{Hettler1998,Kotliar2001,Maier2005}.  Implementations of
these approaches are the dynamical cluster approximation
(DCA)~\cite{Hettler1998,Maier2005} or the cellular
DMFT~\cite{Kotliar2001}. These approaches capture the spin density
wave formation indicated as a band
insulator~\cite{Moukouri2001,Kyung2004} as well as Mott
transitions~\cite{Park2008,Lee2008,Gull2008,Zhang2007,Lee2010(1),Liebsch2009}.
Recently, other implementations including long-range correlations
have been developed by considering a perturbation expansion where
nonlocal contributions are obtained from the two-particle vertex
functions~\cite{Toschi2007,Toschi2011,Rubtsov2008,Li2008,Hafermann2009}.

Recent progress towards a realistic description of correlated systems
is the combination of LDA with DMFT~\cite{Kotliar2006}
(LDA+DMFT). While this approach has proven to be quite successful for
the description of spectral properties of transition metal
oxides~\cite{Aichhorn2009,Pavarini2004,Liebsch2003,Amadon2008,Karolak2010,Kunes2010,Nekrasov2005,Lechermann2006,Nekrasov2006}
and the newly discovered iron-based
superconductors~\cite{Werner2011,Yin2011,Aichhorn2010}, effects
originating from spatial fluctuations remain inconclusive. Attempts to
include short-range spatial fluctuations have been done in the context
of the spin-Peierls system TiOCl -where pairing correlations are
important- within an N$^{th}$ order muffin tin orbital (NMTO) approach
combined with DCA~\cite{Saha-Dasgupta2007} as well as NMTO combined
with a variational cluster approach (VCA)~\cite{Aichhorn2009(1)}.  In
this work, we present an alternative approach where we extend a newly
developed implementation of the LDA+DMFT approach~\cite{Aichhorn2009}
in the context of the full-potential linearized augmented plane wave
(FLAPW)~\cite{Blaha2002} method by including spatial
fluctuations within DCA (LDA+DCA), and we investigate the spectral
properties of SrVO$_3$ as a test case.

The paper is organized as follows: in section~\ref{Formalism}, we
describe our LDA+DCA implementation with a weak-coupling
continuous-time quantum Monte Carlo (CT-QMC)
algorithm~\cite{Rubtsov2005,Assaad2007,Gull2011} for multi-orbital
systems with multiple sites. In section~\ref{Results}, we present
results for SrVO$_3$ within LDA+DCA with a cluster of two sites and
compare them with single-site LDA+DMFT calculations as well as
experimental observations and in section~\ref{Summarize} we summarize
our findings.

\section{THEORETICAL FRAMEWORK}\label{Formalism}

\subsection{LDA+DCA in the APW framework}

In this work, we extend a recent implementation of
LDA+DMFT~\cite{Aichhorn2009} to LDA+DCA which includes short-range
spatial correlations.  We first shortly review the projection
operators within the WIEN2K code~\cite{Blaha2002}.  The local
atomic-like Wannier orbital functions inside an appropriate energy
window $W$ can be expanded over the Bloch basis set as
\begin{equation}
\vert \chi_{{\bf k},m}^{{\alpha},\sigma}\rangle = \sum_{\nu \in W} \langle\psi_{{\bf k},\nu}^{\sigma} \vert 
\chi_{m}^{{\alpha},\sigma}\rangle\vert \psi_{{\bf k},\nu}^{\sigma}\rangle,
\end{equation}
where $\alpha$ indicates the correlated atom, $\nu$ is the band index,
$\sigma$ the spin index, and $m$ the orbital index. Here, $\vert
\psi_{{\bf k}, \nu}^{\sigma}\rangle$ is the Bloch eigenfunction in the
augmented plane wave basis and the correlated orbital $\vert
\chi_{m}^{\alpha,\sigma}\rangle$ is given as $\vert
\chi_{m}^{\alpha,\sigma}\rangle=\vert u_l^{\alpha,\sigma} (E_l)
Y_m^l\rangle$ within the muffin tin sphere, where $E_l$ are chosen
linearization energies, $u_l^{\alpha,\sigma}$ is the radial wave
function, and $Y_m^l$ is the spherical harmonic function.  The
orthonormalized projector operators for the DMFT and DCA
self-consistent equations are calculated by
\begin{equation}
  P_{m,\nu}^{\alpha,\sigma} ({\bf k}) = \sum_{\alpha',m'} \langle u_l^{\alpha',\sigma} (E_l) Y_{m'}^l \vert \psi_{{\bf k}, 
    \nu}^{\sigma}\rangle [O ({\bf k},\sigma)^{-\nicefrac{1}{2}}]_{m,m'}^{\alpha,\alpha'}, 
\end{equation}
where $O ({\bf k},\sigma)_{m,m'}^{\alpha,\alpha'}$ is the overlap function which is given as
\begin{equation}
O ({\bf k},\sigma)_{m,m'}^{\alpha,\alpha'} = \sum_{\nu \in W} \langle\chi_m^{\alpha,\sigma} \vert \psi_{{\bf 
k},\nu}^{\sigma}\rangle\langle\psi_{{\bf k},\nu}^{\sigma} \vert 
\chi_{m'}^{\alpha',\sigma}\rangle. 
\end{equation}

For the LDA+DCA self-consistency procedure, the lattice Green's function is given as
\begin{equation}
G_{\nu,\nu'}^{\sigma} ({\bf K}+\tilde{{\bf k}},i\omega_n) = \frac{1}{i\omega_n + \mu - \epsilon_{{\bf K}+\tilde{{\bf k}},\nu}^{\sigma} 
-\Sigma_{\nu,\nu'}^{\sigma}({\bf K}+\tilde{{\bf k}},i\omega_n)}, \label{eq:latticeGreen}
\end{equation}
where we have defined ${\bf k}$= ${\bf K}+\tilde{{\bf k}}$ with ${\bf
  K}$ being the cluster momenta and $\tilde{{\bf k}}$ running over
each Brillouin zone (BZ) sector. $\omega_n$ is the Matsubara
frequency, $\mu$ is the chemical potential, $\epsilon_{{\bf
    K}+\tilde{{\bf k}},\nu}^{\sigma}$ are the Kohn-Sham (KS)
eigenvalues, and $\Sigma_{\nu,\nu'}^{\sigma}({{\bf K}+\tilde{\bf
    k}},i\omega_n)$ is the lattice self-energy which is calculated as
an expansion of the cluster self-energy over the Bloch basis set:
\begin{equation}\begin{split}
&\Sigma_{\nu,\nu'}^{\sigma} ({\bf K}+\tilde{{\bf k}},i\omega_n) \\ 
&= \sum_{\alpha,m,m'} P_{\nu,m}^{\alpha,\sigma^*} 
({\bf K}+\tilde{{\bf k}}) \Delta \Sigma_{m,m'}^{\sigma,{\rm imp}}({\bf K},i\omega_n) P_{m',\nu'}^{\alpha,\sigma} ({\bf K}+\tilde{{\bf k}}).
\end{split}\end{equation}
%
From the self-energy we need to subtract the contribution to
correlations that is already included in the LDA calculation, commonly
called double counting (DC) correction, 
\begin{equation}
\Delta \Sigma_{m,m'}^{\sigma,{\rm imp}} ({\bf K},i\omega_n) = \Sigma_{m,m'}^{\sigma,{\rm imp}} ({\bf K}, i\omega_n) - \Sigma_{m,m'}^{{\rm dc}},
\end{equation}
where $\Sigma_{m,m'}^{\sigma,{\rm imp}} ({\bf K}, i\omega_n)$ is
calculated by the continuous time Quantum Monte Carlo (CT-QMC) cluster
solver and the Dyson's equation. Calculating the DC correction is not
possible exactly, but some approximate expressions have been
introduced. Here, we use as double counting correction
%
\begin{equation}
\Sigma_{m,m'}^{\sigma,{\rm dc}} = \delta_{m,m'} \Big[U'\Big(N_c-\frac{1}{2}\Big)-J\Big(N_c^{\sigma}-\frac{1}{2}\Big)\Big],
\end{equation}
where $U'= U -2J$ , $U$ is the onsite Coulomb interaction, $J$ is the
Hund's coupling and $N_c$ and $N_c^{\sigma}$ denote the number of
total occupied states and spin-resolved occupied states in the
correlated orbitals, respectively~\cite{Anisimov1997}. The 
%
local
cluster Green's functions are given as
\begin{equation}\begin{split}
G&_{m,m'}^{\sigma,{\rm loc}} ({\bf K},i\omega_n) \\&= \sum_{\tilde{{\bf k}},\nu,\nu'} P_{m,\nu}^{\alpha,\sigma} 
({\bf K}+\tilde{\bf k}) G_{\nu,\nu'}^{\sigma} ({\bf K}+\tilde{{\bf k}},i\omega_n) P_{\nu',m'}^{\alpha,{\sigma^*}'}({\bf K}+\tilde{\bf k}),
\label{eq:dcaself}
\end{split}\end{equation}
where the summation over $\tilde{{\bf k}}$ is calculated in each
Brillouin zone sector.  The LDA+DCA self-consistency condition states
that this local cluster Green's functions, Eq. (\ref{eq:dcaself}),
have to be equal to the impurity Green's functions as calculated by
CT-QMC. The DMFT update of the Weiss field is given by the Dyson's
equation as 
\begin{equation}
\big[G_{m,m'}^{\sigma,0} ({\bf K},i\omega_n)\big]^{-1} = \Sigma_{m,m'}^{\sigma,{\rm imp}} ({\bf K},i\omega_n) + \big[G_{m,m'}^{\sigma,{\rm loc}} ({\bf K},i\omega_n)\big]^{-1}.
\label{eq:dyson}
\end{equation}

\subsection{Many-body interactions and CT-QMC algorithm}

In order to describe the electronic behavior of SrVO$_3$ 
%
one has to
consider
the multiorbital Hubbard Hamiltonian where the interaction term is
given by:
\begin{equation}\begin{split}
H_{I} =& U \sum_m n_{m\uparrow}  n_{m\downarrow} + 
\sum_{m<n,\sigma}\big[U' n_{m\sigma} n_{n\bar{\sigma}} \\ 
&+ (U'-J)n_{m\sigma}n_{n \sigma}  - J' c_{m\sigma}^{\dagger}
c_{m\bar{\sigma}} c_{n\bar{\sigma}}^{\dagger} c_{n\sigma} \\
&- J' c_{m\sigma}^{\dagger} c^{\dagger}_{m\bar{\sigma}} c_{n\sigma} c_{n\bar{\sigma}} \big],
\label{eq:hami}
\end{split}\end{equation}
and $m$,$n$ denote the $t_{2g}$ orbitals. In order to solve this model
we employ a weak-coupling CT-QMC algorithm.  While the weak-coupling
CT-QMC algorithm can easily treat a multiple number of sites in the
cluster, it is difficult to deal with the full rotationally invariant
form of the interaction Hamiltonian due to the fermionic sign problem,
in contrast to the strong-coupling CT-QMC
algorithm~\cite{Werner2006,Werner2007}.  Therefore, in what follows we
shall consider a simplified Hubbard model where the spin-flip and
pair-hopping terms in Eq. (\ref{eq:hami}) are neglected 
($J'=0$).

The main idea of the weak-coupling CT-QMC method is to divide the
total action $S$ into an unperturbed term $S_0$ and the interaction
term $I$ which is expanded in a Taylor series. The partition function
is rewritten as
\begin{equation} \label{eq:ctqmc}
\begin{aligned}
{\cal Z} = \sum_{k}Z_{0}\frac{(-I)^{k}}{k!}\int d\tau_{1}\cdots d\tau_{k} \int{\cal D}[c,\bar{c}]\\
{\langle{n_{l_1\uparrow}(\tau_{1})n_{l'_1\downarrow}(\tau_{1})\cdots 
n_{l_k\uparrow}(\tau_{k})n_{l'_k\downarrow}(\tau_{k})}\rangle},
\end{aligned}
\end{equation}
where
${\langle{n_{l_1\uparrow}(\tau_{1})n_{l'_1\downarrow}(\tau_{1})\cdots
    n_{l_k\uparrow}(\tau_{k})n_{l'_k\downarrow}(\tau_{k})}\rangle}$ is
determined by the non-interacting Green's function and Wick's theorem,
$k$ is the perturbation order, $Z_{0}={\rm Tr}(Te^{-S_0})$ corresponds
to the unperturbed term and $l$, $l'$ and $\tau_k$ are randomly
sampled. $I$ is given as
\begin{equation}
I=\bar {U}\beta N M(2M-1),
\end{equation}
where $\beta$ is the inverse temperature, $N$ and $M$ are the number
of sites and the number of orbitals in the cluster, respectively, and
$\bar{U}$ is one of $U$, $U'$ or $U'-J$ depending on the operators
considered in the random walk in the average $\langle\dots\rangle$ in
Eq.~\eqref{eq:ctqmc}.
The impurity Green's functions are calculated by
numerically averaging Eq. (\ref{eq:ctqmc}).

\section{RESULTS}\label{Results}

SrVO$_3$, which is thought to be a prototypical paramagnetic
correlated metal with intermediate electron-electron interactions, has
served in the past as a testing ground for numerous newly developed
LDA+DMFT
approaches\cite{Aichhorn2009,Pavarini2004,Liebsch2003,Amadon2008,Karolak2010,Kunes2010,Nekrasov2006}.
In SrVO$_3$, the V $3d$-orbitals are split by the crystal field into
triply degenerate $t_{2g}$ and doubly degenerate $e_g$ states. The LDA
calculations show that the degenerate $t_{2g}$ states of V form bands
crossing the Fermi level which are well separated from the $e_g$
bands.

For our calculations on SrVO$_3$, we chose the energy window $W$ from
-1.35~eV to 2.0~eV for the $t_{2g}$ orbitals which can then be
effectively described by the degenerate three-orbital Hubbard model in
Eq. (\ref{eq:hami}).  We first reproduced the results of LDA+DMFT
from Ref.~\onlinecite{Aichhorn2009}, considering a temperature
$T=0.1$~eV and the same Coulomb interaction $U=4.0$~eV, and Hund's
rule coupling $J=0.65$~eV.  In a next step, we extend the LDA+DMFT
solution to LDA+DCA with two sites in the cluster $N=2$.  
Within the DCA method, $N=2$ implies that we  have
two BZ sectors and the self-energies in the BZ sectors are
constant.

\begin{figure}
\includegraphics[width=0.48\textwidth]{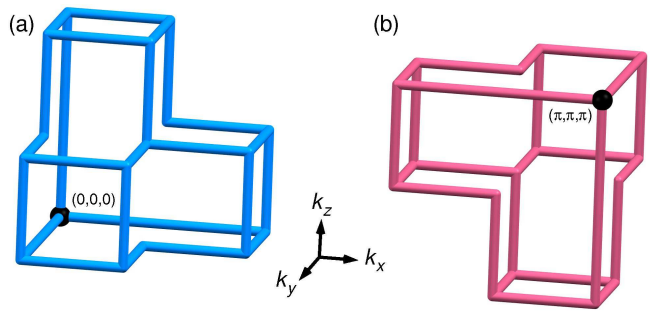}
\caption{
  (Color online) One quadrant of the Brillouin zone of SrVO$_3$. (a) and (b) represent the
  ${\bf K}_{0}=(0,0,0)$ and ${\bf K}_{1}=(\pi,\pi,\pi)$ Brillouin zone
  sectors, respectively. Other quadrants follow from
  symmetry.} \label{fig:DCABZ}
\end{figure}

The BZ of SrVO$_3$ has cubic symmetry and the self-energies in the
cluster momenta ${\bf K}_{0}=(0,0,0)$ and ${\bf K}_{1}=(\pi,\pi,\pi)$
are calculated in the BZ sectors shown in Figs.~\ref{fig:DCABZ}(a) and
(b).  In real space, the on-site and nearest neighbor-site Green's
functions are $G_{R=0} (i\omega_n) = \frac{1}{2} (G_{\bf K_0}
(i\omega_n) + G_{\bf K_1} (i\omega_n))$ and $G_{R=1} (i\omega_n) =
\frac{1}{2} (G_{\bf K_0} (i\omega_n) - G_{\bf K_1} (i\omega_n))$,
respectively.  Here, the DCA formalism with $N=2$ for cubic lattice
has been clearly presented in Ref.~\onlinecite{Lin2009}. Both on-site and nearest neighbor-site Green's
functions are inserted into the CT-QMC impurity solver, and the
LDA+DCA self-consistency is satisfied by Eqs. (\ref{eq:dcaself}) and
(\ref{eq:dyson}).

\begin{figure}
\includegraphics[angle=270,width=0.43\textwidth]{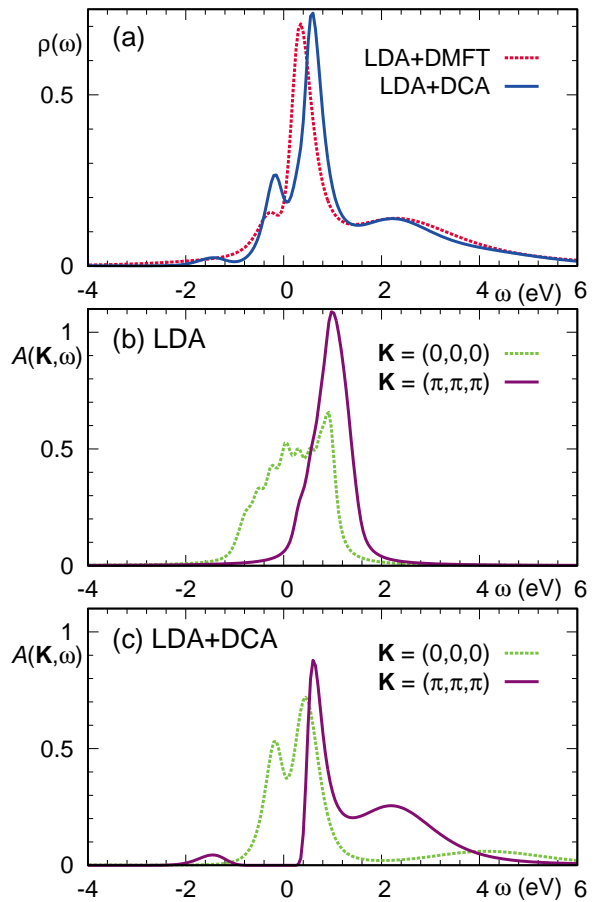}
\caption{(Color online) (a) The density of states $\rho(\omega)$ of SrVO$_3$
  calculated within the LDA+DMFT and LDA+DCA approaches, with 
  $U=4.0$~eV, $J=0.65$~eV and $T=0.1$~eV.  The density of states
  $\rho(\omega)$ for LDA+DCA is calculated by $\rho(\omega)=
  \frac{1}{2} (A({\bf K}_{0},\omega)+A({\bf K}_{1},\omega))$, with
  ${\bf K}_{0}=(0,0,0)$ and ${\bf K}_{1}=(\pi , \pi, \pi)$ sectors.
  (b) and (c) The spectral functions $A({\bf K},\omega)$ obtained from
  LDA and LDA+DCA for the ${\bf K}_{0}$ and ${\bf K}_{1}$ sectors by
  Eq. (\ref{eq:dcaself}), respectively.  All the density of states and
  spectral functions are normalized to one.}\label{fig:DOS}
\end{figure}

Fig.~\ref{fig:DOS}(a) shows the density of states $\rho(\omega)$ of the vanadium $t_{2g}$ orbitals obtained within the LDA+DMFT,
$\rho(\omega)$= $A(\omega)$ and LDA+DCA, $\rho(\omega)= \frac{1}{2}
(A({\bf K}_{0},\omega)+A({\bf K}_{1},\omega))$. Here, the spectral function $A({\bf K},\omega)$
is given as 
\begin{equation}
 A({\bf K},\omega) = -\frac{1}{\pi} {\bf Im} G_{\bf K} (\omega),
\end{equation}
and an analytical continuation of the impurity Green's functions is performed through
a maximum entropy method. Our LDA+DMFT results obtained by both the weak-coupling
CT-QMC~\cite{Rubtsov2005,Assaad2007} as well as the strong-coupling
CT-QMC algorithms from the ALPS code~\cite{Bauer2011} agree with
former LDA+DMFT
calculations~\cite{Aichhorn2009,Pavarini2004,Amadon2008,Karolak2010,Kunes2010}.
In Figs.~\ref{fig:DOS}(b) and (c) we present the spectral functions
for the ${\bf K}_{0}=(0,0,0)$ and ${\bf K}_{1}=(\pi , \pi, \pi)$
sectors within LDA and LDA+DCA.  The new features obtained in LDA+DCA
are a broad peak around 1.5 eV and a coherent peak around 0.2 eV below
$E_{\rm F}$. LDA results (see Fig.~\ref{fig:DOS}(b)) as well as most
former LDA+DMFT results (see also Fig.~\ref{fig:DOS} (a)) don't
exhibit neither a broad peak nor a clear coherent peak below $E_{\rm F}$.
Recent angle resolved photoemission (ARPES)
experiments~\cite{Yoshida2010}, have observed, in fact, a broad peak
around 1.5 eV and a coherent peak around 0.4 eV below $E_{\rm F}$ (Fig. 3(b) in Ref. ~\onlinecite{Yoshida2010}).  
We suggest that the better agreement of the LDA+DCA with ARPES
observations is a consequence of the inclusion of short-range spatial
correlations. Fig.~\ref{fig:DOS}(c) shows that the coherent and broad
peaks below the Fermi level are caused by the distinct spectral
weights in the ${\bf K}_{0}=(0,0,0)$ and ${\bf K}_{1}=(\pi , \pi,
\pi)$ sectors.  These two sectors also show respectively metallic and
band insulating behavior reminiscent of the LDA results in these
sectors (Fig~\ref{fig:DOS}(b) and Ref. ~\onlinecite{Wadati2006}).

\begin{figure}
\includegraphics[angle=270,width=0.52\textwidth]{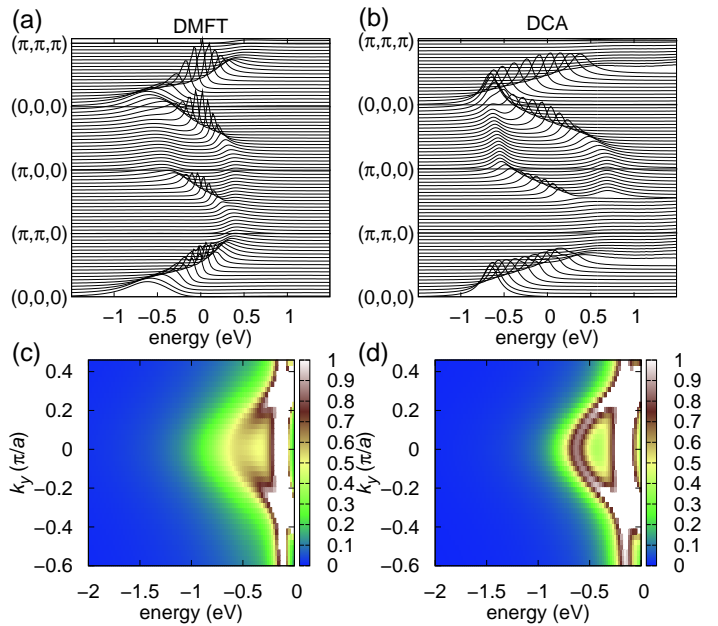}
\caption{(Color online) (a) and (b) Spectral functions obtained within (a) LDA+DMFT and (b) LDA+DCA for
  the vanadium $t_{2g}$ bands. (c) and (d) Spectral functions in the regime between
$k_y=-0.6\frac{\pi}{a}$ and $0.4\frac{\pi}{a}$ at $k_x=0.0$ and $k_z=0.32\frac{\pi}{a}$
within (c) LDA+DMFT and (d) LDA+DCA for the vanadium $t_{2g}$ bands.}\label{fig:edc}
\end{figure}

In Figs.~\ref{fig:edc} (a) and ~\ref{fig:edc} (b), we show the momentum resolved spectral functions
calculated from Eq. (\ref{eq:latticeGreen}) and
Eq. (\ref{eq:dcaself}).  The analytical continuation of the
self-energy $\Sigma ({\bf K},i\omega_n)$ is performed by the maximum
entropy approach with subtraction of the Hartree-Fock
term~\cite{Chen2011}. In view of the ill-posed problem of the
analytical continuation of the self-energy~\cite{Wang2009}, we also
compared the DOS obtained from integration of the spectral functions
with those in Fig.~\ref{fig:DOS}~(a) and found a reasonable
agreement. We also compare LDA+DMFT to LDA+DCA results. 
One can observe some redistribution of momentum
resolved spectral weight between the LDA+DMFT and LDA+DCA results.
 In Figs.~\ref{fig:edc} (c) and ~\ref{fig:edc} (d) we plot the
 LDA+DMFT and LDA+DCA spectral functions respectively,  in the region between
$k_y=-0.6\frac{\pi}{a}$ and $0.4\frac{\pi}{a}$ at $k_x=0.0$
 and $k_z=0.32\frac{\pi}{a}$  ($a$ is the lattice constant) 
in order to directly compare our calculations
to the ARPES results (Fig. 1 (a) of Ref.~\cite{Yoshida2010}).
In agreement with ARPES experiments, both LDA+DMFT and LDA+DCA 
 show dispersive features around -0.7 though they are more pronounced in
the LDA+DCA calculations. Also, the LDA+DCA calculations reproduce
the small peak observed around -0.2eV.
These results account for the renormalization of
the bands due to electronic correlations. 


Finally, our estimation of the mass enhancement is $m^*/m \approx 1.7 \pm
0.3$ within LDA+DMFT and $1.6 \pm 0.5$ within LDA+DCA at ${\bf K}_{0}=(0,0,0)$. These values
are obtained from
\begin{equation}
m^*/m \approx 1 - \frac{\partial \rm{Im}\Sigma (i\omega_n)}{\partial \omega}\Big\vert_{\omega \rightarrow 0^{+}},
\end{equation}
where the derivative is extracted by fitting a third-order polynomial
to the lowest four Matsubara frequencies~\cite{Mravlje2011}. 
Note that our LDA+DCA estimates give a slightly smaller mass enhancement than
LDA+DMFT estimates with a larger error bar. Both sets of results
are in accordance with ARPES estimates of $m^*/m
\approx 1.8 \pm 0.2$~\cite{Yoshida2005}.

\section{CONCLUSIONS}\label{Summarize}
In conclusion, we have presented an implementation of the LDA+DCA
method within the linear augmented plane-wave framework.  We have
compared our benchmark results on {\svo}, which is modeled in terms of
a three-band Hubbard Hamiltonian, with earlier LDA+DMFT calculations
as well as experimental data. Since the LDA+DCA approach considers
both local quantum as well as short-range spatial fluctuations, it
offers a more complete description of correlated materials compared to
the LDA+DMFT approach, where only local quantum fluctuations are taken
into account.

Unlike the LDA+DMFT, the LDA+DCA approach reproduces both coherent and
broad peaks for SrVO$_3$ below the Fermi level, observed in angle integrated photoemission
experiments.  The analysis of the spectral functions at ${\bf
  K}_{0}=(0,0,0)$ and ${\bf K}_{1}=(\pi,\pi,\pi)$ reveals the source
of these peaks.  While the broad peak is due to the spectral function
in the ${\bf K}_{0}=(0,0,0)$ sector, the coherent peak has its origin
in the spectral function at the ${\bf K}_{1}=(\pi,\pi,\pi)$ sector.
We also observe a metallic and a band insulating state at the ${\bf
  K}_{0}=(0,0,0)$ and ${\bf K}_{1}=(\pi,\pi,\pi)$ sectors, also
present in the LDA results.

In summary, we believe that the presented LDA+DCA approach is very
promising and can be applied to a large variety of multiorbital
correlated compounds at different fillings.

\section{Acknowledgements}
We would like to thank Y.-Z. Zhang and C. Gros for useful discussions and we
gratefully acknowledge financial support from the Deutsche
Forschungsgemeinschaft through grants FOR 1346 (H.L.) and SPP 1458
(J.F.) and from the Helmholtz Association through grant HA216/EMMI.
M.A. acknowledges support from the Austrian Science Fund, project
F4103, and hospitality at Goethe-Universit\"at Frankfurt.

\end{document}